\documentclass{article}
\usepackage{graphicx} 
\usepackage{natbib}
\usepackage{amsmath}
\usepackage{hyperref}

\title{Quantifying the evolving topical structure of science across journals, countries, regions, and research domains}
\author{Carlos Oscar Sorzano\\
Natl. Center of Biotechnology (CNB-CSIC), Madrid, Spain\\
\texttt{coss@cnb.csic.es}}\date{February 2026}

\begin{document}

\maketitle

\begin{abstract}
Timely and comparable indicators of the evolving structure of science are
increasingly needed for research policy and strategic planning. We present a reproducible and scalable framework for quantifying the topical prevalence and recent dynamics of scientific activity using open scholarly metadata from OpenAlex. The approach combines a unified topic ontology with simple trend estimators derived from short time series, enabling consistent comparisons across journals, countries, regions, and domain-focused corpora.

We illustrate the methodology through representative case studies spanning generalist journals, national output, metropolitan research ecosystems, and structural biology. Across these examples, the framework captures both system-level normalization effects and fine-grained specialization patterns. Because the pipeline is fully general and based on open data, it can be readily extended to continuous, multi-scale monitoring of the scientific landscape.

The proposed methodology provides a compact and interpretable quantitative layer that can complement expert assessment in science policy, research evaluation, and strategic decision-making.
\end{abstract}


\section{Introduction}

Strategic decisions in science policy, institutional planning, and research investment increasingly rely on quantitative assessments of the evolving structure of science \citep{Fortunato2018}. Governments must prioritize research areas under constrained budgets, regions seek to specialize effectively within national and international ecosystems, and companies aim to locate expertise where it is most concentrated. At the same time, high-visibility journals such as \emph{Science} and \emph{Nature} continue to shape perceptions of the scientific frontier \citep{lariviere2010, Brembs2013}. Despite the importance of these decisions, the information available to policymakers and strategic planners often remains fragmented, delayed, or influenced by local visibility and expert perception. There is therefore a growing need for reproducible, large-scale indicators that provide an objective and timely view of how scientific activity is distributed and evolving across topics, geographic scales, and research domains.

Over the past decades, bibliometrics and science mapping have produced a rich toolbox for analyzing the organization of research. Citation-based maps \citep{Boyack2010}, keyword co-occurrence analyses \citep{Callon1991}, and expert-curated classification systems \citep{Borner2012, Waltman2012} have substantially improved our understanding of the structure of scientific knowledge. However, many of these approaches face practical limitations when used for continuous, policy-oriented monitoring. Citation-network methods are computationally demanding and difficult to update frequently; keyword-based analyses can be unstable, sensitive to vocabulary drift, and require active use in bibliographic databases; expert panels, while indispensable, do not scale easily and may suffer from bounded visibility. In addition, cross-scale comparisons—such as contrasting journals, countries, regions, and domain-specific corpora within a unified framework—remain methodologically challenging.

The emergence of the science of science (SciSci) has reinforced the value of large-scale quantitative analyses for understanding the dynamics of research systems \citep{Fortunato2018}. Prior work in this area has uncovered robust regularities in citation behavior \cite{Wang2013e}, collaboration networks \citep{Newman2001}, and career trajectories \citep{Sinatra2016}. However, much of the literature has emphasized either universal mechanisms or complex network and generative models, while comparatively less attention has been devoted to lightweight, transparent indicators that can support continuous monitoring across multiple geographic and thematic scales. Bridging this gap requires operational frameworks that translate SciSci's conceptual advances into scalable observatories of the scientific landscape.

Recent advances in open scholarly data infrastructures—most notably OpenAlex \citep{Priem2022}—now make it possible to revisit these questions from a new perspective. These platforms provide topic-level annotations for millions of publications together with rich affiliation metadata, enabling reproducible analyses that were previously difficult to conduct at a global scale. 

In this work, we introduce a reproducible framework for quantifying and comparing the evolving topical structure of scientific production across multiple aggregation levels. Using a unified topic ontology (the one from OpenAlex) and short longitudinal windows, we estimate both the prevalence of research topics and their recent growth trends within (i) generalist high-impact journals, (ii) national research outputs, (iii) metropolitan ecosystems, and (iv) specific research domains such as structural biology. This unified perspective enables systematic comparisons, including identifying areas where a country lags behind the global editorial frontier, detecting regional specialization patterns, and recognizing topics that are underrepresented in high-visibility venues relative to their broader research activity.

Our goal is not to replace expert judgment but to complement it with scalable, data-driven situational awareness \citep{Hicks2015}. By providing transparent, reproducible indicators derived from open data, the proposed approach can help reduce blind spots in science policy discussions, support evidence-based specialization strategies, and inform decisions ranging from national funding priorities to the geographic placement of industrial R\&D activities \citep{Wilsdon2015}. More broadly, the framework contributes to the development of continuous, open observatories of the scientific landscape that can track structural changes in research systems in near-real time. 

Recent work has demonstrated the feasibility of constructing global overlay maps of science using open infrastructures such as OpenAlex, enabling visualization of how the publication profiles of authors or institutions are positioned within a citation-based map of the scientific landscape \citep{Haunschild2024}. These approaches provide valuable geometric views of the science system by embedding concepts according to direct citation relations and overlaying unit-specific publication counts. However, their primary focus is spatial positioning within a fixed global topology, and they rely on relatively heavy network construction and visualization pipelines. In contrast, the present work aims to develop lightweight, fully reproducible indicators that quantify topical prevalence and recent dynamics across multiple aggregation levels. Rather than emphasizing geometric embedding, our framework prioritizes temporal sensitivity, cross-scale comparability, and policy-oriented interpretability, enabling continuous monitoring of how research portfolios evolve across journals, countries, regions, and scientific domains.

The remainder of the paper presents the methodological framework, describes its implementation using OpenAlex data, and illustrates its utility through comparative analyses of generalist journals, countries, metropolitan research regions, and selected scientific domains.

\section{Methods}

The complete implementation of the methodology described in this work is publicly available as a reproducible Jupyter notebook at \url{https://github.com/cossorzano/COSS_DataAnalysis_notebooks/blob/main/Notebooks/researchTopicsTrends.ipynb}. The notebook contains the full data retrieval pipeline, processing steps, and analysis code required to reproduce all results reported in this study. Below, we summarize the main methodological components.

\subsection{Data retrieval and aggregation}

All analyses were based on the OpenAlex open scholarly metadata platform, accessed through its public REST API. The data collection pipeline was implemented in Python and designed to operate in a fully reproducible, restartable manner, with automatic retry mechanisms and cursor-based pagination.

The framework supports multi-scale aggregation of scientific output across four types of units: (i) generalist journals (e.g., \emph{Science}, \emph{Nature}), (ii) countries based on author affiliations, (iii) metropolitan regions defined through institutional geolocation, and (iv) domain-focused journal sets (e.g., structural biology venues).

Journal-level datasets were obtained by querying the OpenAlex source identifier associated with each venue. Country-level datasets were constructed by selecting works for which at least one author was affiliated with an institution in the target country (ISO-3166 alpha-2 code). Regional datasets were built in two stages: first, institutions located in the target city were identified using text search combined with country filtering and post-validation of the reported city; second, works associated with each institution were retrieved year by year. To prevent dominance by very large institutions and to stabilize regional comparisons, the number of works retrieved per institution and year was capped at a predefined maximum. Duplicate work arising from multi-institution collaborations was removed while preserving the order.

Because the number of retrieved works can be large, the extraction process was implemented incrementally. Intermediate dictionaries were periodically serialized to disk using atomic writes, allowing safe interruption and resumption while avoiding reprocessing of previously analyzed records.

\subsection{Topic extraction and counting}

For each retrieved work, we obtained from OpenAlex the publication year, the primary topic, the full list of assigned topics with confidence scores, and the associated field and subfield classifications. Identifiers were normalized to their short OpenAlex form and stored alongside lookup tables that map identifiers to display names.

Yearly occurrence counts were then computed for each topic, field, and subfield. Counts were based on per-work presence: a topic contributed one count per work regardless of its confidence score, and fields and subfields were counted once per work. When necessary, missing year entries were densified with zeros to produce complete time series across the five-year study window.

In addition to yearly counts, the overall prevalence for each entity was computed as the fraction of works in the study window that contained the corresponding topic, field, or subfield. These quantities provide complementary information on the relative importance of research areas independent of their temporal trend.

\subsection{Trend estimation}

To quantify recent dynamics, we fitted a linear model to the five-year count trajectory of each entity (a similar approach was followed in \citet{Bornmann2015}). Let $y_t$ denote the yearly count. Years were mapped to a centered index
\[
x \in \{-2,-1,0,1,2\},
\]
and the model
\[
y = \beta_0 + \beta_1 x
\]
was estimated using ordinary least squares. The slope $\beta_1$ was interpreted as an indicator of recent growth ($\beta_1>0$) or decline ($\beta_1<0$), while the intercept provides the central level of activity.

All computed indicators, including yearly counts, fitted coefficients, and prevalence values, were exported in tabular form for downstream comparative analyses across journals, countries, regions, and research domains.

\section{Results}

The analyses presented below are intended primarily as illustrative applications of the proposed framework rather than as an exhaustive mapping of the scientific landscape. Our objective is to demonstrate how a unified, reproducible pipeline can quantify topical prevalence and temporal dynamics across multiple aggregation levels, including generalist journals, national outputs, regional ecosystems, and domain-focused corpora. The methodology is fully general and can be applied to any collection of venues, countries, regions, or thematic subsets available in OpenAlex or similar scholarly databases.

In the following subsections, we therefore focus on a set of representative
case studies that highlight different facets of the approach: the editorial
profile of \emph{Science} and \emph{Nature}, the national research structure
of Spain, the regional contrast between Madrid and Barcelona, and a
domain-specific analysis of structural biology. Together, these examples
illustrate the framework's capacity to reveal both broad systemic patterns
and fine-grained signals of specialization. Complete result tables generated
from the referenced notebook for each of these analyses are provided as
Supplementary Material.

\subsection{Field dynamics in \emph{Science} and \emph{Nature}}

At the field level, the largest intercept values ($>500$) are observed for \emph{Medicine}, \emph{Biochemistry, Genetics and Molecular Biology}, \emph{Environmental Science}, \emph{Social Sciences}, \emph{Physics and Astronomy}, and \emph{Engineering}. These areas, therefore, constitute the dominant thematic backbone of the combined \emph{Science}–\emph{Nature} corpus during the study window. Among them, \emph{Medicine} stands out by a wide margin, with an intercept exceeding 1700, followed by molecular and environmental sciences. When examining temporal dynamics, the most pronounced trends (defined by $|\beta_1|\ge 15$) are overwhelmingly negative. The steepest decline is observed in \emph{Medicine} ($\beta_1=-134.2$), with substantial additional decreases in \emph{Environmental Science}, \emph{Physics and Astronomy}, \emph{Social Sciences}, and \emph{Biochemistry, Genetics and Molecular Biology}. Consistent with these field-level patterns, the total number of papers in the combined corpus decreases from 11{,}517 in 2021 to 8{,}810 in 2025, corresponding to an overall reduction of approximately $24\%$.

To place field-specific dynamics in the context of the overall contraction of the corpus (approximately $-24\%$ between 2021 and 2025), we examined the scaled ratio $4\beta_1/\beta_0$, which provides an approximate estimate of the relative change over the full five-year window. Most high-volume backbone fields exhibit declines of comparable magnitude to the global trend. For example, \emph{Medicine} and \emph{Physics and Astronomy} show relative decreases on the order of $-30\%$, while \emph{Environmental Science} closely tracks the system-wide average.

More revealing, however, is the presence of a group of fields that display substantially greater resilience. In particular, \emph{Biochemistry, Genetics and Molecular Biology}, \emph{Engineering}, and \emph{Agricultural and Biological Sciences} show markedly smaller relative contractions, indicating stable representation within \emph{Science} and \emph{Nature} despite the overall decrease in publication volume. An even stronger signal emerges for \emph{Decision Sciences}, \emph{Computer Science}, \emph{Neuroscience}, \emph{Psychology}, \emph{Chemistry}, and \emph{Arts and Humanities}, several of which exhibit near-zero or positive relative trends. 

Taken together, these patterns indicate that the post-2021 contraction is not uniform across the disciplinary spectrum, but is accompanied by a relative shift toward computational, quantitative, and cross-disciplinary areas.

\subsection{Field dynamics in Spain}

The Spanish corpus exhibits a markedly different global behavior from that observed in the combined \emph{Science}–\emph{Nature} dataset. The total number of retrieved Spanish papers remains essentially constant across the study window (2021–2025), fluctuating within a narrow band around approximately 10{,}900 publications per year. However, this apparent stability must be interpreted in light of the data collection strategy. Because the retrieval pipeline imposes a maximum number of works per year, the Spanish dataset represents a capped subset of the national output, biased toward the most cited publications. As a consequence, the flat global trajectory primarily indicates saturation of the retrieval threshold rather than the absence of underlying changes in total national production.

Despite this ceiling effect at the aggregate level, the internal structure of the Spanish research portfolio reveals clear and informative rebalancing. The largest intercept values correspond to \emph{Medicine}, \emph{Engineering}, \emph{Biochemistry, Genetics and Molecular Biology}, \emph{Environmental Science}, and \emph{Materials Science}, which together define the dominant thematic backbone of Spanish scientific activity. Compared with the profiles of \emph{Science} and \emph{Nature}, Spain places greater emphasis on engineering, materials, and applied physical sciences.

Relative dynamics, assessed through the scaled ratio $4\beta_1/\beta_0$, highlight a coherent expansion in several technology- and materials-oriented domains. In particular, \emph{Chemistry}, \emph{Materials Science}, \emph{Engineering}, \emph{Energy}, and \emph{Chemical Engineering} exhibit clear positive trends, with \emph{Physics and Astronomy} also showing moderate growth. These signals point to a strengthening of the physical sciences and engineering base within the Spanish high-impact publication segment.

By contrast, the strongest relative contractions are observed in \emph{Psychology}, \emph{Economics, Econometrics and Finance}, \emph{Business, Management and Accounting}, and \emph{Social Sciences}.

Overall, while the capped Spanish corpus does not exhibit a global decline in volume, it shows a clear internal redistribution toward engineering, materials, energy, and chemistry-related domains, accompanied by a relative weakening in several socio-economic and behavioral fields.

To further contextualize the Spanish research profile, we compared the fractional share of each field and subfield in Spain with the corresponding share in the combined \emph{Science}–\emph{Nature} corpus. The ratio Spain/Journals therefore quantifies the relative representation of each area in the national portfolio with respect to the editorial frontier defined by these high-impact journals. Fields and subfields with Spain/Journals $<0.5$ can be interpreted as areas in which Spain is comparatively underrepresented within the \emph{Science}–\emph{Nature} ecosystem. Notable examples include \emph{Arts and Humanities} at the field level, and several subfields such as \emph{Anthropology}, \emph{Reproductive Medicine}, \emph{Paleontology}, \emph{Political Science and International Relations}, and \emph{Archeology}. In these areas, the Spanish share in elite journals is substantially smaller than would be expected from its overall research portfolio. From a strategic perspective, such gaps may indicate opportunities to increase international visibility, although the appropriate policy response depends on broader national priorities and mission considerations.

At the opposite extreme, fields and subfields with Spain/Journals $>2$ correspond to areas that are relatively strong within the Spanish system, but comparatively less represented in \emph{Science} and \emph{Nature}. Prominent examples include subfields related to \emph{Control and Systems Engineering}, \emph{Computer Science Applications}, \emph{Computer Networks and Communications}, \emph{General Energy}, and \emph{Building and Construction}, as well as several clinically oriented biomedical areas. In these domains, Spain has substantially greater internal weight than is reflected in generalist high-impact journals. This pattern suggests that evaluation frameworks relying heavily on \emph{Science} and \emph{Nature} publications may systematically undervalue nationally important research areas.

Taken together, the comparative analysis reveals a structured asymmetry between the Spanish research portfolio and the topical profile of \emph{Science} and \emph{Nature}. Rather than indicating simple underperformance, these differences point to a combination of editorial selectivity effects and genuine national specialization patterns. The Spain/Journals ratio, therefore, provides a quantitative lens for identifying both potential visibility gaps and areas that may be important at the country level but that are not well represented by elite generalist journals.

\subsection{Regional specialization: Madrid versus Barcelona}

To investigate subnational structure within the Spanish research system, we compared the relative field composition of Madrid and Barcelona. Fields were considered strongly differentiated when the relative weight of one region exceeded that of the other by at least a factor of 1.4.

A coherent Madrid-leaning cluster emerges in the physical and engineering sciences. The strongest asymmetries are observed in \emph{Materials Science}, \emph{Energy},
\emph{Chemical Engineering}, \emph{Engineering}, \emph{Physics and Astronomy}, and \emph{Chemistry}, with \emph{Agricultural and Biological Sciences} also showing a clear Madrid advantage. These results indicate that Madrid concentrates a disproportionately large share of activity in materials, energy, and applied physical domains.

In contrast, Barcelona shows relative specialization in the life and biomedical sciences. The most prominent Barcelona-leaning areas include \emph{Biochemistry, Genetics and Molecular Biology}, \emph{Immunology and Microbiology}, \emph{Neuroscience}, and \emph{Medicine}. This pattern is consistent with a stronger biomedical and molecular research profile in the Barcelona ecosystem.

Overall, the Madrid–Barcelona comparison reveals a clear and structured division of thematic emphasis within Spain, with Madrid more strongly oriented toward engineering, energy, and the physical sciences, and Barcelona toward biomedical and life sciences research. The magnitude of several differences indicates that these are not marginal fluctuations but reflect deep structural specialization across the two major metropolitan research hubs.

\subsection{Domain analysis: Structural Biology}

We next focus on a domain-defined corpus representative of structural biology. The dataset was constructed by aggregating publications from a curated set of specialized journals widely recognized in the field: \emph{Journal of Structural Biology} (JSB), \emph{Current Opinion in Structural Biology} (COSB), \emph{Nature Structural \& Molecular Biology} (NSMB), \emph{Acta Crystallographica Section D}, and \emph{Acta Crystallographica Section F}. For this corpus, we summarize composition and dynamics at three hierarchical levels (fields, subfields, and topics), using the fitted intercept as a measure of overall prevalence during the study window and the fitted slope $\beta_1$ as an indicator of recent change.

\paragraph{Fields.} At the field level, the prevalence of the domain is strongly concentrated in \emph{Biochemistry, Genetics and Molecular Biology} (intercept $\approx 539.4$), followed at a much lower level by \emph{Medicine} ($\approx 150.2$) and \emph{Materials Science} ($\approx 147.8$). A second tier includes \emph{Chemistry} ($\approx 46.4$), \emph{Agricultural and Biological Sciences} ($\approx 37.2$), \emph{Immunology and Microbiology} ($\approx 34.6$), \emph{Physics and Astronomy} ($\approx 33.6$), and \emph{Computer Science} ($\approx 32.4$), highlighting a measurable multidisciplinary footprint around a predominantly molecular-bioscience core. Considering only strong trends ($|\beta_1|\ge 3$), three fields stand out: a sharp decline in \emph{Materials Science} ($\beta_1=-20.6$), a moderate decline in \emph{Chemistry} ($\beta_1=-3.9$), and a clear positive trend in \emph{Computer Science} ($\beta_1=4.2$), consistent with increasing computationalization of structural biology workflows.

\paragraph{Subfields.} At the subfield level, the dominant component is \emph{Molecular Biology} (intercept $\approx 465.8$), followed by a substantial \emph{Materials Chemistry} contribution ($\approx 119.6$). Other highly prevalent subfields include \emph{Structural Biology} ($\approx 54.6$), \emph{Cell Biology} ($\approx 48.6$), and \emph{Genetics} ($\approx 41.0$), again reflecting a molecular-life-science nucleus with adjacent methodological and materials-related layers. Using the $|\beta_1|\ge 3$ criterion, the strongest subfield-level signals are concentrated in two directions: a marked negative trend for \emph{Materials Chemistry} ($\beta_1=-20.2$) and a positive trend for \emph{Computational Theory and Mathematics} ($\beta_1=4.1$). This combination suggests a domain-level reweighting away from materials-oriented subfields and toward computational foundations.

\paragraph{Topics.} At the finest resolution, the most prevalent topics include \emph{RNA and protein synthesis mechanisms} (intercept $\approx 109.2$), \emph{Enzyme Structure and Function} ($\approx 103.2$), and \emph{Protein Structure and Dynamics} ($\approx 94.2$), followed by \emph{RNA Research and Splicing} ($\approx 59.8$), \emph{Genomics and Chromatin Dynamics} ($\approx 53.8$), and \emph{Advanced Electron Microscopy Techniques and Applications} ($\approx 54.6$). Among topics with strong temporal trends ($|\beta_1|\ge 3$), the steepest declines are observed for \emph{Enzyme Structure and Function} ($\beta_1=-21.6$) and \emph{Protein Structure and Dynamics} ($\beta_1=-10.4$), together with additional negative shifts in \emph{Biochemical and Molecular Research} ($\beta_1=-4.1$), \emph{Bacterial Genetics and Biotechnology} ($\beta_1=-3.3$), and \emph{Glycosylation and Glycoproteins Research} ($\beta_1=-3.2$). In contrast, several topics show strong positive dynamics, notably \emph{RNA modifications and cancer} ($\beta_1=6.0$), \emph{RNA Research and Splicing} ($\beta_1=4.7$), \emph{Computational Drug Discovery Methods} ($\beta_1=4.0$), and \emph{Genomics and Chromatin Dynamics} ($\beta_1=3.3$). Overall, the topic-level signals indicate that, within structural biology, RNA-centric and computational/virtual-screening related directions are expanding, while several standard protein/enzyme-structure topics show substantial contraction during the same period.

\section{Discussion}

The multi-scale analyses presented here reveal a coherent but nuanced picture of the recent evolution of the scientific landscape. By applying a unified framework across generalist journals, national outputs, and regional ecosystems, along with a domain-focused corpus, we show that the same quantitative pipeline can reveal both global normalization effects and fine-grained specialization patterns.

A first robust signal emerges from the combined \emph{Science}–\emph{Nature} corpus, which exhibits a broad contraction of approximately $24\%$ over the 2021--2025 window. Importantly, however, the contraction is not uniform. Several large backbone areas, including \emph{Biochemistry, Genetics and  Molecular Biology} and \emph{Engineering}, show comparatively mild declines, while computationally oriented fields display relative resilience. This pattern is consistent with a gradual rebalancing of editorial space toward data-intensive and cross-disciplinary research.

In contrast, the Spanish corpus does not display an aggregate decline. However, this apparent stability must be interpreted cautiously, as the national dataset is subject to an upper retrieval limit. Under these conditions, the time series effectively tracks the most visible segment of the Spanish research portfolio rather than total national output. Within this high-impact slice, a clear internal redistribution is observed, characterized by strengthening in engineering, materials, energy, and chemistry-related domains, alongside relative weakening in several socio-economic and behavioral areas. The divergence between the editorial frontier and the national profile highlights the importance of considering both perspectives when assessing research performance.

The comparison between Madrid and Barcelona further illustrates the framework's ability to capture meaningful subnational structures. The results reveal a pronounced thematic complementarity between the two metropolitan ecosystems. Madrid shows a strong tilt toward engineering, energy, materials, and the physical sciences, whereas Barcelona places greater emphasis on biomedical and life sciences. The magnitude and coherence of these differences indicate that they reflect genuine structural specialization rather than random fluctuations, supporting the view that the Spanish research system benefits from functional regional differentiation.

The domain-focused analysis of structural biology provides an additional layer of insight. While the field remains overwhelmingly anchored in molecular bioscience, internal dynamics indicate a measurable shift in methodological emphasis. Materials-related components show notable contraction, whereas computational science and RNA-centered topics display clear positive trends. These observations are consistent with the broader technological evolution of structural biology toward increasingly computation-intensive, integrative, and data-driven workflows.

Several methodological considerations should be kept in mind when interpreting these results. First, the use of full counting assigns equal weight to all contributing entities within each work, which may amplify the influence of large collaborations and disregards citation counts (one advantage of doing so is that citation patterns vary across fields and require specialized techniques to be homogenized \citep{Hutchins2016}). Second, the capped retrieval strategy in the national and regional analyses emphasizes the most visible portion of the output and may attenuate global volume changes. Third, topic assignment in large-scale scholarly databases is inherently probabilistic and may introduce classification noise at fine resolution. Despite these caveats, the consistency of the observed multi-level patterns supports the robustness of the main conclusions.

From a strategic perspective, the proposed indicators provide a compact and interpretable lens for evidence-informed decision making. Differences between national portfolios and the topical profile of elite journals may signal either visibility gaps or legitimate specialization choices. Likewise, regional asymmetries can inform smart specialization strategies, and domain-level analyses can reveal emerging methodological transitions within fields. Importantly, the framework is not intended to prescribe policy but to provide a quantitative situational awareness layer that complements expert judgment.

More broadly, the present work illustrates the feasibility of constructing reproducible, scalable observatories of the scientific landscape using open scholarly metadata. Because the pipeline is fully general, it can be readily extended to other countries, regions, domains, or time windows, and can support continuous monitoring of structural changes in science. As open metadata infrastructures continue to mature, such approaches may become an increasingly valuable component of the evidence base for science policy and research strategy.

\section{Conclusions}

We have introduced a unified and reproducible framework to quantify the topical structure and recent dynamics of scientific activity across multiple aggregation levels, including generalist journals, national outputs, metropolitan ecosystems, and domain-focused corpora. By leveraging OpenAlex metadata and simple, interpretable trend indicators, the proposed approach enables consistent multi-scale comparisons of the research landscape.

The case studies presented here demonstrate the framework's ability to capture both system-level normalization effects and fine-grained specialization patterns across geographic and thematic dimensions. While the empirical examples are illustrative, they confirm that lightweight indicators derived from open metadata can provide meaningful situational awareness of the evolving structure of science.

The primary contribution of this work is therefore methodological. Because the pipeline is fully general, transparent, and based on open data, it can be readily extended to other countries, regions, journals, or scientific domains and can support continuous monitoring of research systems over time. As open scholarly infrastructures continue to mature, approaches of this type may become an increasingly valuable complement to expert assessment in science policy, research evaluation, and strategic planning.

\section*{Competing interests}

The authors have no relevant financial or non-financial interests to disclose.


\end{document}